\documentstyle[prl,aps]{revtex}
\begin{document}
\title{Mass as a relativistic quantum observable}
\author{Marc-Thierry Jaekel$^a$ and Serge Reynaud$^b$}
\address{$(a)$ Laboratoire de Physique Th\'{e}orique de l'Ecole Normale
Sup\'{e}rieure, Universit\'{e} de Paris-Sud, \\
Centre National de la Recherche Scientifique,
24 rue Lhomond, F-75231 Paris Cedex 05, France\\
$(b)$ Laboratoire Kastler Brossel, Universit\'{e} Pierre et Marie Curie,
Ecole Normale Sup\'{e}rieure, \\
Centre National de la Recherche Scientifique,
4 place Jussieu, F-75252 Paris Cedex 05, France}
\date{Europhys. Lett. 38 (1997) 1}
\maketitle

\begin{abstract}
A field state containing photons propagating in different directions has a
non vanishing mass which is a quantum observable. We interpret the shift of
this mass under transformations to accelerated frames as defining space-time
observables canonically conjugated to energy-momentum observables.
Shifts of quantum observables differ from the predictions of
classical relativity theory in the presence of a non vanishing spin. 
In particular, quantum redshift of energy-momentum is affected by spin. 
Shifts of position and energy-momentum observables 
however obey simple universal rules derived
from invariance of canonical commutators.


\end{abstract}

Since the birth of classical mechanics, the two fundamental concepts of
space and mass have essentially been regarded as physical notions defined
simultaneously but independently of one another. While space is the arena
where motion takes place, mass is a measure of inertia associated with a
given moving object. General theory of relativity introduces a dynamical
connection between space and mass since inertia and gravitation are
interpreted in terms of geometrical properties of space-time \cite
{Einstein16}. We will argue in the present letter that mass and space-time
have also to be considered as closely connected in quantum physics, even in
the absence of gravitation. When set in the framework of quantum theory,
seminal arguments developed by Einstein in the early years of relativity
theory indeed entail that mass should not be considered as a constant, but
rather as a relativistic quantum observable the properties of which allow to
define positions in space-time.

In accordance with the standard relativistic prescription \cite{Einstein0506}, 
the mass $M$ associated with a physical system will be defined as the
Lorentz invariant built on the energy-momentum variables $P_\mu $ 
\begin{equation}
M=\sqrt{\eta ^{\mu \nu }P_\mu P_\nu }=\sqrt{P_\mu P^\mu }  \label{defM}
\end{equation}
Throughout the letter, the Minkowski tensor $\eta ^{\mu \nu }$ is used to
raise or lower indices. Definition (\ref{defM}) leads to attribute inertia
to any kind of energy and implies that mass is a dynamical quantity rather
than a constant parameter. For instance, the mass of an
atom changes when it emits or absorbs a photon. It is also known that energy
and momentum are shifted under transformations to accelerated frames. The
proportionality of these shifts versus acceleration and position measured
along the direction of acceleration reveals the equivalence between uniform
acceleration and constant gravity \cite{Einstein07}. Mass, as defined by
equation (\ref{defM}), also undergoes a redshift (or a blueshift) under
transformations to accelerated frames.

These issues raise intriguing questions in the context of quantum theory.
Since energy and momentum are quantum observables,
redshifts have to be written in terms of quantum positions which
necessarily differ from parameters used to map space-time.
Hence, relativistic transformations of physical 
observables, like the Einstein redshift law, may no longer be
given their usual derivation from classical relativity theory. 
Moreover, covariance properties are associated with conventional choices 
of coordinate maps, and the universal form of the redshift laws must be
given a new interpretation in terms of quantum observables. Dimensional
relations between mass and time connect this question to the similar issue
of universality of relativistic transformations of space-time observables.
The velocity of light $c$ and the Planck constant $\hbar $ are fundamental
constants, so that mass scale is expected to vary as the inverse of time
scale under frame transformations \cite{DSH}. Clearly, any formulation where
the status of mass is that of a constant classical parameter does not have
the ability to solve these questions. A few attempts have been made to
develop theoretical approaches where mass is allowed to vary or described as
a quantum variable but they rely on specific features which are added to the
standard formalism and do not appear to be universally admitted \cite{GB}.

The approach developed in the present letter uses the fact that, although
photons are massless, any field state containing photons propagating in at
least two different directions corresponds to a non vanishing mass (\ref
{defM}). As an illustration, one may consider a state built on two
counterpropagating photons which corresponds to a vanishing momentum and
therefore has a non vanishing mass equal to its energy. We will thus be able
to study a mass treated as a quantum observable in the simple framework of
free electromagnetic theory. We will then show that the mass shift under
transformations to accelerated frames may be written in terms of space-time
observables canonically conjugated to energy-momentum observables. Symmetry
of electromagnetism under Lorentz transformations plays a prime role in the
deduction of relativitistic effects. Here, we will use conformal invariance
of electromagnetism which allows to deal with uniformly accelerated frames 
in classical \cite{BCFRW} as well as quantum physics \cite{BFH83}.

Conformal symmetry is basically described by conformal algebra, that is the
set of commutators of its generators. Commutators of generators $\left[
\Delta _a,\Delta _b\right] $, and more generally of all observables, will be
written under the form 
\begin{equation}
\left( \Delta _a,\Delta _b\right) \equiv \frac 1{i\hbar }\left[ \Delta
_a,\Delta _b\right]
\end{equation}
In the following, we will often use the Jacobi identity  
\begin{equation}
\left( \left( \Delta _a,\Delta _b\right) ,\Delta _c\right) =\left( \Delta
_a,\left( \Delta _b,\Delta _c\right) \right) -\left( \Delta _b,\left( \Delta
_a,\Delta _c\right) \right)  \label{Jacobi}
\end{equation}
Conformal generators are integrals of the energy-momentum density of the
quantum fields which are preserved under field propagation \cite{Itz}. They
are defined in such a manner that they vanish in vacuum, 
in consistency with conformal invariance of vacuum \cite{QSO95}. 
Corresponding respectively to
translations (energy-momentum $P_\nu $), rotations (angular momentum and
Lorentz boosts $J_{\nu \rho }$), dilatation ($D$) and conformal
transformations to uniformly accelerated frames ($C_\nu $), the 
generators obey the following commutation relations 
\begin{eqnarray}
&&\left( P_\mu ,P_\nu \right) =0\qquad \left( J_{\mu \nu },P_\rho \right)
=\eta _{\nu \rho }P_\mu -\eta _{\mu \rho }P_\nu  \nonumber \\
&&\left( J_{\mu \nu },J_{\rho \sigma }\right) =\eta _{\nu \rho }J_{\mu
\sigma }+\eta _{\mu \sigma }J_{\nu \rho }-\eta _{\mu \rho }J_{\nu \sigma
}-\eta _{\nu \sigma }J_{\mu \rho }  \nonumber \\
&&\left( D,P_\mu \right) =P_\mu \qquad \left( D,J_{\mu \nu }\right) =0 
\nonumber \\
&&\left( P_\mu ,C_\nu \right) =-2\eta _{\mu \nu }D-2J_{\mu \nu }  \nonumber
\\
&&\left( J_{\mu \nu },C_\rho \right) =\eta _{\nu \rho }C_\mu -\eta _{\mu
\rho }C_\nu  \nonumber \\
&&\left( D,C_\mu \right) =-C_\mu \qquad \left( C_\mu ,C_\nu \right) =0
\label{ConfAlg}
\end{eqnarray}
Equations (\ref{ConfAlg}) follow from the commutation rules of quantum
fields. They describe the commutators of the generators considered as
quantum observables as well as the frame transformations of the generators
considered as relativistic observables. The $P_\mu {}$'s are commuting
components of a Lorentz vector. The $C_\mu {}$'s obey the same property. The
conformal weights, determined by commutation relations with $D$, are
opposite for $C_\mu {}$'s and $P_\mu {}$'s. The $J_{\mu \nu }{}$'s form a
Lorentz tensor with a vanishing conformal weight. Finally, the commutators
between the $P_\mu {}$'s and $C_\nu {}$'s describe the shifts of
energy-momentum under transformations to accelerated frames. As quantum
analogs of redshift laws, they play an important role in the following
discussions.

Focusing our interest on the mass observable $M$ defined in equation (\ref
{defM}), we write its shifts under different frame transformations 
\begin{eqnarray}
\left( P_\mu ,M\right)  &=&\left( J_{\mu \nu },M\right) =0\qquad \left(
D,M\right) =M  \nonumber \\
\left( C_\mu ,M\right)  &=&2\left( \eta _{\mu \rho }D-J_{\mu \rho }\right)
\cdot \frac{P^\rho }M  \label{DM}
\end{eqnarray}
We have taken care of non commutativity of observables by introducing a
symmetrised product represented by the ``$~\cdot ~$'' symbol. As expected,
mass shift vanishes for Poincar\'{e} transformations and is
proportional to $M$ for dilatation. For transformations to
accelerated frames, mass shift exhibits a state dependence that allows to
define space-time observables. To emphasize this key point, we consider the
general transformation to an accelerated frame 
\begin{equation}
\Delta =\frac{a^\mu }2C_\mu   \label{AF}
\end{equation}
where the classical numbers $a^\mu $ represent accelerations along the
various space-time directions. The mass shift $\left( \Delta ,M\right) $
corresponding to such a transformation may be written  
\begin{equation}
\left( \Delta ,M\right) =a^\mu M\cdot X_\mu    \qquad
X_\mu \equiv \left( \eta _{\mu \rho }D-J_{\mu \rho }\right) 
\cdot \frac{P^\rho }{M^2}  \label{defX}
\end{equation}
Shifts under Poincar\'{e} transformations and dilatation
of the observables $X_\mu $ defined in this manner are derived
from Jacobi identity (\ref{Jacobi}) and conformal algebra (\ref{ConfAlg})    
\begin{equation}
\left( P_\mu ,X_\nu \right) =-\eta _{\mu \nu }\qquad \left( D,X_\mu
\right) =-X_\mu   \qquad
\left( J_{\mu \nu },X_\rho \right) =\eta _{\nu \rho }X_\mu -\eta _{\mu
\rho }X_\nu   \label{PX}
\end{equation}
They are identical to the shifts of coordinate parameters under
corresponding map changes. The observables $X_\mu $ are conserved quantities
since they have been built on conformal generators. These remarks suggest
that these quantum observables have to be interpreted as the positions 
of some event in space-time.
In the specific case of a state consisting in two light pulses 
originating from a given point \cite{PLA96}, observables $X_\mu $ 
may effectively be identified with the space-time positions 
of the coincidence event. 
The same interpretation still holds in the general case of  
an arbitrary number of photons propagating along any direction, 
provided the position of the coincidence event is understood
as an average over
the energy-momentum distribution associated with the quantum field state. 
In the spirit of the discussion presented in the introduction,
relations (\ref{defX}) reveal an intimate connection between mass and
space-time. The mass shift under frame transformation (\ref{AF}) is indeed
proportional to the mass $M$ itself, to the acceleration $a^\mu $ and to the
position $X_\mu $ measured along the direction of acceleration. This is
exactly the form expected for the potential energy of a mass in a constant
gravitational field so that equation (\ref{defX}) may be interpreted as the
redshift of mass written in terms of quantum
positions. This law hence constitutes a statement of equivalence between
acceleration and gravity, valid in the quantum domain.

The first of equations (\ref{PX}) also means that observables 
$X_\mu $ are conjugated to energy-momentum operators.
Canonical commutation relations are thus
embodied in conformal algebra (\ref{ConfAlg}). Although they have been
derived directly from standard quantum formalism, these equations
conflict statements which are often claimed to be unavoidable
consequences of this formalism \cite{UW89}. As a matter of fact, equations (%
\ref{defX},\ref{PX}) contain the definition of a time operator $X_0$ besides
that of space operators. Furthermore, this operator enters an energy-time
canonical commutation relation which has the same form as momentum-space
relations while the whole set of relations satisfies Lorentz invariance. In
accordance with the definition of conformal generators (\ref{ConfAlg}), time
as well as space operators are defined only for field states orthogonal to
vacuum, so that difficulties related to hermiticity in the definition of
phases are bypassed \cite{Phases}. We may stress again
that the time operator $X_0$ is a localisation observable,
that is precisely the date associated with an event, 
and therefore differs from any kind of evolution parameter \cite{PLA96}. 
The shifts studied in the present letter do not represent an effect of
evolution but rather an effect of frame transformations.
In particular, canonical commutators describe the shifts of space-time
observables under space-time translations. With these precisions kept in
mind, relations (\ref{defX},\ref{PX}) contain a quantum definition for a
time operator \cite{Jammer74}.

The previous derivations still hold in the presence of spin which is
however known to play a crucial role in the problem of localisability of
quantum fields \cite{PNWF}. This is clearly illustrated by the evaluation
of commutators of space-time components, using
Jacobi identity (\ref{Jacobi}), conformal algebra (\ref{ConfAlg}) 
and canonical commutators (\ref{PX})
\begin{eqnarray}
&&\left( M\cdot X_\mu ,M\cdot X_\nu \right) =\frac 14\left( \left( C_\mu
,M\right) ,\left( C_\nu ,M\right) \right) =J_{\mu \nu } \nonumber \\
&&M^2 \cdot \left( X_\mu ,X_\nu \right) = J_{\mu \nu }-
\left( P_\mu \cdot X_\nu -P_\nu \cdot X_\mu \right)
\equiv S_{\mu \nu } \label{XX}
\end{eqnarray}
These commutators are given by expressions
$S_{\mu \nu }$ defined as the difference between angular momentum 
$J_{\mu \nu }$ and orbital angular momentum 
built on energy-momentum and space-time observables.
This internal angular momentum is directly related to the 
Pauli-Lubanski vector $S^\mu$ which is the standard covariant generalisation 
of spin in relativistic kinematics \cite{Itz,Rohrlich}
\begin{equation}
S_{\mu \nu } = \epsilon _{\mu \nu \rho \sigma } S^\rho 
\frac{P^\sigma}{M}  \qquad
S^\mu = - \frac 12\epsilon ^{\mu \nu \rho \sigma }J_{\nu \rho } 
\frac{P_\sigma}{M}  \label{deffS}
\end{equation}
where $\epsilon _{\mu \nu \lambda \rho }$ is the completely antisymmetric
Lorentz tensor of rank $4$. 

Equations (\ref{XX})
entail that commuting positions may be defined only in the specific case of a
vanishing spin. Since spin is a basic ingredient of quantum physics, the
preceding relation shows that notions inherited from classical differential
geometry cannot be applied without modification to quantum observables. As a
further illustration, the redshift $\left(
\Delta ,P_\nu \right) $ of energy-momentum under the frame transformation 
(\ref{AF}) is evaluated from (\ref{ConfAlg}) as 
\begin{equation}
\left( \Delta ,P_\nu \right) =a ^\mu \left( \eta _{\mu \nu }D-P_\mu
\cdot X_\nu +P_\nu \cdot X_\mu -S_{\mu \nu }\right) \qquad D=P_\rho \cdot
X^\rho   \label{CP}
\end{equation}
The expression of $D$ which appears in equation (\ref{CP}) is a direct
consequence of the definition (\ref{defX}) of position observables.
The quantum redshift law (\ref{CP}) differs from the classical one, since the
redshift of energy-momentum now depends not only on positions but also on
spin observables. Einstein's law must be regarded as a classical
approximation valid in the limiting case where spin contributions are
negligible. In all cases however, mass shift keeps a classical form 
(\ref{defX}) which is not affected by spin contributions. This is
related to the fact that these contributions are orthogonal 
to energy-momentum in equation (\ref{CP}).

The redshift law (\ref{CP}) has a universal form
dictated by conformal algebra, although this form differs from the 
classical one.
Shifts of positions $X_\mu $ take the form predicted by classical
relativity for Poincar\'{e} transformations and dilatation (see (\ref{PX}))
but not for transformations to accelerated frames which, already for a
vanishing spin, mix positions with energy-momentum \cite{PLA96,PRL96}. The
derivation of these shifts in the presence of spin lies outside the scope of
the present letter. We may however give simple expressions which 
provide interesting insights concerning the question of
universality of relativistic transformations. To this aim, we note that the
canonical commutators $\left( P_\mu ,X_\nu \right) $ are pure numbers which
are invariant under all frame transformations.
Invariance of canonical commutators corresponds in fact to the
statement that the Planck constant is constant like the velocity of light 
\cite{DSH}, not only for Poincar\'{e} transformations but also for 
conformal transformations to accelerated frames. Since
$\left( P_\mu ,X_\nu \right) $ commutes with the generator $\Delta $
of such transformations,
Jacobi identity (\ref{Jacobi}) leads to the following relation
\begin{equation}
\left( \left( \Delta ,X_\nu \right) ,P_\mu \right)  =\left( \left( \Delta
,P_\mu \right) ,X_\nu \right)   \label{invComm}
\end{equation}
To discuss the significance of this relation, we first remind
that $\left( \Delta ,X_\nu \right) $ is the shift of position, so that $%
\left( \left( \Delta ,X_\nu \right) ,P_\mu \right) $ is the variation of
this shift under a translation. This expression is thus a quantum analog of
the matrix $\frac{\partial \delta x^\nu }{\partial x^\mu }$ which is used in
classical differential geometry to describe the change of a vector
under the infinitesimal coordinate deformation 
$\delta x^\nu (x^\mu )$. The second double commutator $\left( \left( \Delta
,P_\mu \right) ,X_\nu \right) $ appearing in (\ref{invComm}) is similar to
the first one with the roles of position and energy-momentum interchanged.
The presence of a non vanishing spin affects the expression (\ref
{CP}) of the redshift $\left( \Delta ,P_\mu \right) $ as well as the
evaluation of commutators with position components (see (\ref{XX})). 
In this context, equation (\ref{invComm}) exhibits a non
trivial relation between shifts of quantum observables associated with
position and energy-momentum which is still valid in the presence of a non
vanishing spin. 
Furthermore, an explicit evaluation of the double commutators
shows that the expressions appearing in equation (\ref{invComm})
have a classical form which is not affected by spin
\begin{equation}
\left( \left( \Delta ,X_\nu \right) ,P_\mu \right)  
=\left( \left( \Delta ,P_\mu \right) ,X_\nu \right)    
= - \eta _{\mu \nu } a ^\rho X_\rho - a _\mu X_\nu + a _\nu X_\mu    
\label{covRules}
\end{equation}
In the particular case $\mu =\nu =0$, 
the two expressions connected by equation (\ref{invComm}) represent
respectively the shift of a clock rate under acceleration and the commutator
with time of the redshift of energy. 
Equation (\ref{covRules}) 
identifies their common expression in terms of quantum positions 
with the prediction of classical relativity although
the time shift $\left( \Delta ,X_0\right) $ 
and the redshift $\left( \Delta ,P_0\right) $ both differ from
classical predictions.

To sum up our results, we have defined a mass observable for a quantum field
state and derived a position in space-time from the redshift of this mass, in
accordance with Einstein redshift law. Space-time observables defined in
such a manner have been found to be canonically conjugated to
energy-momentum observables. We have also derived transformation laws which
differ from the predictions of classical relativity theory but keep a
universal form dictated by conformal algebra. Commutators of
position components as well as redshifts of energy-momentum are affected
by spin. Covariance rules which reflect the universality of predictions of
classical relativity theory have been reformulated as statements of
invariance of canonical commutators under frame transformations.

Strictly speaking, all these results have been established only for specific
quantum systems, namely free electromagnetic fields with photons propagating
in at least two different directions. Arguing that their range of interest
is restricted to such specific systems would however lead to considerable
difficulties concerning the consistency between relativistic and quantum
properties as well as the known universality of these properties. Since a
consistent theory including general relativity and quantum theory is still
lacking, it is worth discussing basic concepts at the interplay of quantum
theory, inertia and gravity. The results of the present letter plead for a
theoretical frame where mass and space-time positions would be treated as
relativistic quantum observables with their basic properties embodied in
symmetries rather than in classical covariance rules \cite{Norton93}. They
already demonstrate that equivalence between uniform acceleration and
constant gravity fits perfectly well in such a frame. Reanalysing the
questions raised by gravitational physics should probably lead to reconsider
the role played in quantum theory by basic geometrical concepts \cite
{Connes95}.

\end{document}